\documentclass[final,3p,times]{elsarticle}

\usepackage[T1]{fontenc}       
\usepackage{newtxtext,newtxmath}
\usepackage[utf8]{inputenc}

\usepackage{amssymb}
\usepackage{amsmath}
\usepackage{multicol}
\usepackage[english]{babel}
\usepackage{color}
\usepackage{bm}
\usepackage{natbib}

\usepackage{booktabs}

\journal{Physica A}
\begin{document}

\begin{frontmatter}

\title{Influence of Functional Group on the Self Assembly of Diamondoids: A Molecular Dynamics Study}

\author[label1]{Sonam Phuntsho}

\address[label1]{Department of Physical Science, Sherubtse College, Royal University of Bhutan,
            42007 Kanglung,
            Trashigang,
            Bhutan}

\begin{abstract}
This study investigates the molecular-level self-assembly behavior of seven functionalized diamondoids, examining how diverse substituents influence structural organization, thermal stability, and aggregate morphology. Using a combination of density functional theory for initial geometry optimization and molecular dynamics simulations, we explored radial distribution functions, thermal fragmentation temperatures, and radii of gyration for each system. Our results reveal that hydrogen-bonding and polar functional groups (e.g., amino, hydroxy) foster well-defined, ordered assemblies, while bulkier or less interactive substituents (e.g., phenyl, methoxy) lead to more open, amorphous aggregates. Thermal stability strongly depends on substituent chemistry: complex, bulky groups or heteroatom-rich functionalities confer enhanced resistance to fragmentation at high temperatures, whereas simpler groups destabilize the assembly at lower temperatures. Radii of gyration further show that substituent size and polarity can fine-tune cluster compactness. These findings provide critical insights for designing diamondoid-based nanomaterials with tailored structural properties, thermal endurance, and functional performance in advanced technological applications.

\end{abstract}


\end{frontmatter}

\section{Introduction}
Diamondoids are cage-like, hydrogen-terminated hydrocarbons that possess a diamond-like lattice structure at the nanoscale \cite{1}. Their unique structural properties, such as high thermal stability and rigidity, make them attractive candidates for various applications in nanotechnology, medicine, and materials science \cite{2, 3}. Functionalization of diamondoids, attaching specific chemical groups to the diamondoid core can significantly alter their physical and chemical properties, enabling tailored functionalities for specific applications \cite{4, 5}. For instance, functionalized diamondoids have been explored for use in drug delivery systems, electronic devices, and as building blocks for advanced materials \cite{6,7,8}.

The self-assembly of functionalized diamondoids holds promise for the bottom-up fabrication of nanostructured materials \cite{9,10}. Understanding the molecular dynamics of these processes is crucial for designing advanced materials with desired properties. Previous studies have explored the self-assembly behavior of simple diamondoids and their derivatives. Ciesielski et al. \cite{11} examined molecular self-assembly on graphite and its dependency on molecular concentration, revealing ordered 2D structures. Noh et al. \cite{12} studied poly(norbornene)-based block copolymers functionalized with adamantane and their transformation into micellar structures due to host-guest interactions.

Despite these advancements, comprehensive molecular dynamics studies focusing on the self-assembly of specifically functionalized diamondoids remain limited. Yeung et al. \cite{13} highlighted the need for more systematic investigations into functionalized diamondoids for advanced materials. Rai et al. \cite{14} conducted molecular dynamics simulations of to investigate the structural properties of self-assembled monolayers. The findings indicated non-linear thermal effects on molecular tilt angles, pointing to gaps in understanding these system

Analyses such as radial distribution functions (RDF) provide insights into the structural organization of molecular assemblies \cite{15,16}. RDF analyses help elucidate intermolecular distances and interactions, which are critical in understanding self-assembly mechanisms \cite{17}. Thermal stability analyses are essential for determining the feasibility of these materials in practical applications, especially under varying temperature conditions \cite{18}. The radius of gyration is another critical parameter that reflects the overall size and compactness of molecular assemblies \cite{19}.

This study involves molecular dynamics of self-assembly in seven functionalized diamondoids: 3,5,7-tetraamino \\ adamantane, tetracyanoadamantane, 1,3,5,7-tetrahydroxyadamantane, 2,2-dimethoxyadamantane, 1,3,5,7-tetraphenyl \\adamantane, 2,8,9-trioxa-1-phosphadamantane, and methyl 1-adamantanesulfonate. These compounds were selected due to their varied functional groups, which can influence their self-assembly behavior and resultant material properties \cite{20,21}.

The primary objectives of this study are to analyze the radial distribution functions of the assembled structures to understand intermolecular interactions, assess the thermal stability of the assemblies under different temperature conditions, and evaluate the radius of gyration to determine the size and compactness of the molecular assemblies. Understanding these aspects will provide valuable insights into how functional groups influence the self-assembly process of diamondoids, which is significant for the design of novel nanomaterials
\cite{22,23,24}.

\section{Methodology}
We explored the self-assembly behavior of seven functionalized diamondoids: 1,3,5,7-tetraaminoadamantane, tetracyanoadamantane, 1,3,5,7-tetrahydroxyadamantane, 2,2-dimethoxyadamantane, 1,3,5,7-tetraphenyladamantane, 2,8,9-trioxa-1-phosphadamantane, and methyl 1-adamantanesulfonate. For each diamondoid, molecular dynamics (MD) simulations were conducted with 125 identical molecules under vacuum conditions.

The initial molecular geometries were optimized using density functional theory (DFT) with the B3LYP exchange-correlation functional \cite{25,26} and the cc-pVDZ basis set \cite{27} implemented in the Gaussian 09 software package \cite{28}. Natural Bond Orbital (NBO) analysis \cite{29} was performed to obtain atomic partial charges required for the MD simulations.
MD simulations were carried out using the OPLS-AA (Optimized Potentials for Liquid Simulations All-Atom) force field \cite{30}, suitable for modeling organic molecules and their interactions. Simulations were performed using the LAMMPS software package \cite{31}.
Simulation parameters included the use of real units and a timestep of 1.0 femtoseconds (fs).  Non-bonded interactions were modeled using the Lennard-Jones potential with a cutoff distance of 10.0\,\AA. Long-range Coulombic interactions were handled using the particle-particle particle-mesh (PPPM) method with an accuracy of $1.0 \times 10^{-4}$. Neighbor lists were updated every timestep with modifications to efficiently handle large systems. Periodic boundary conditions were applied in all three dimensions.
\begin{equation*}
U
= \sum_{\text{bonds}} \frac{1}{2} k_b (r - r_0)^2 
+ \sum_{\text{angles}} \frac{1}{2} k_\theta (\theta - \theta_0)^2 
+ \sum_{\text{dihedrals}} \sum_{n} \frac{V_n}{2} [1 + \cos(n\varphi - \gamma)] 
+ \sum_{i<j} \left[4\epsilon_{ij}\bigg\{\left(\frac{\sigma_{ij}}{r_{ij}}\right)^{12} 
- \left(\frac{\sigma_{ij}}{r_{ij}}\right)^{6}\bigg\} 
+ \frac{q_i q_j}{4 \pi \varepsilon_0 r_{ij}}\right]
\end{equation*}

The systems were initialized by assigning velocities corresponding to a temperature of 290\,K using the Maxwell-Boltzmann distribution. An NVT ensemble was employed using the Nosé-Hoover thermostat \cite{32} to maintain a constant temperature of 290\,K. Each simulation was run until all the diamondoids self assembled. 
To assess thermal stability, the temperature was gradually increased using the Nosé-Hoover thermostat at 1 K/ps. 

\section{Results and Discussion}
\subsection{Structural Analysis via Radial Distribution Functions (RDF)}
\begin{figure}
	\centering
	\includegraphics[width=.4\columnwidth]{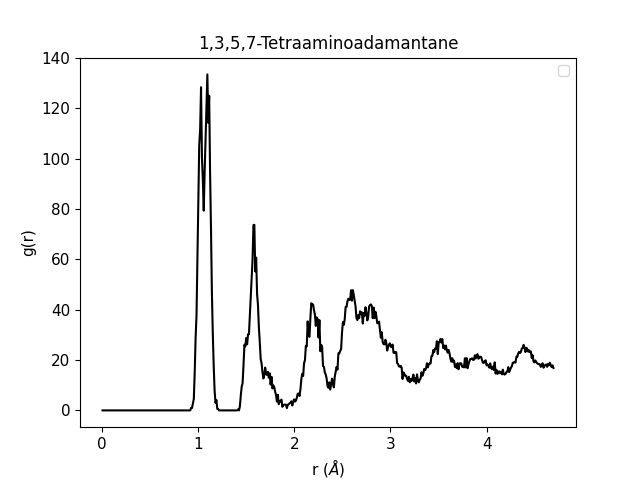}
	\includegraphics[width=.4\columnwidth]{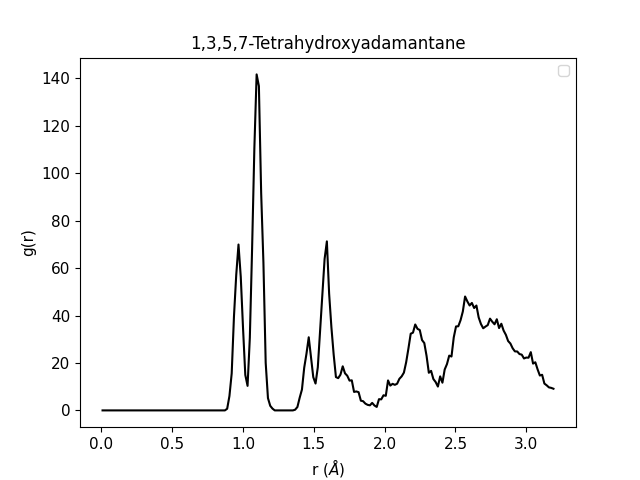}
    \includegraphics[width=.4\columnwidth]{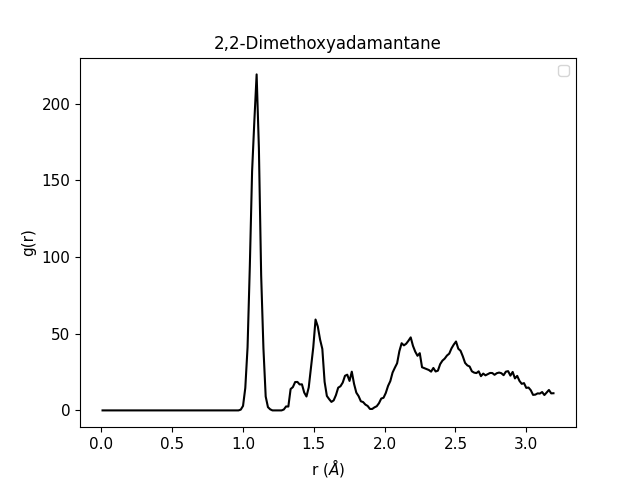}
    \includegraphics[width=.4\columnwidth]{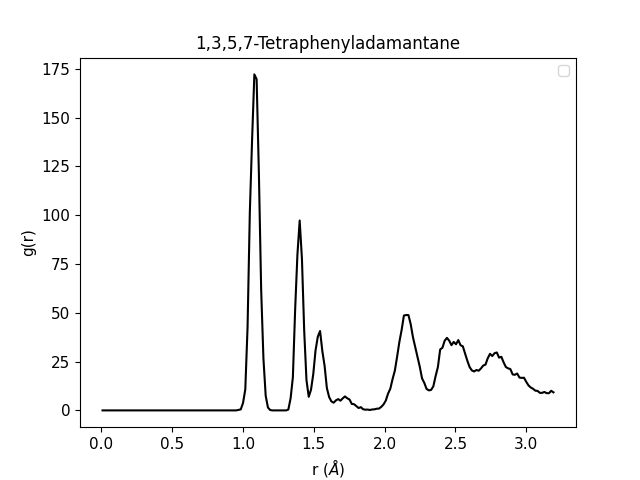}
    \includegraphics[width=.4\columnwidth]{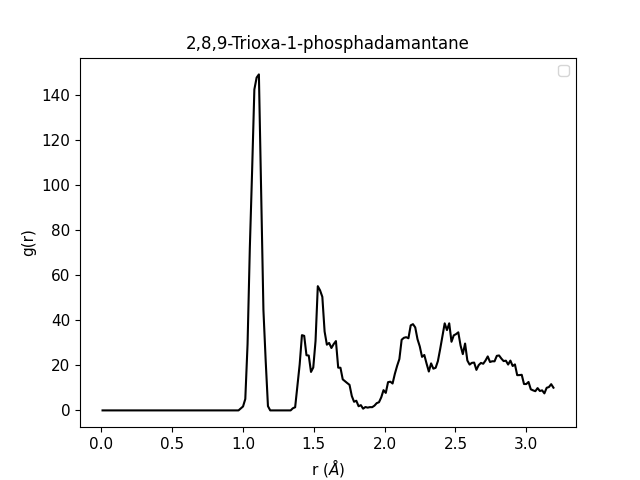}
    \includegraphics[width=.4\columnwidth]{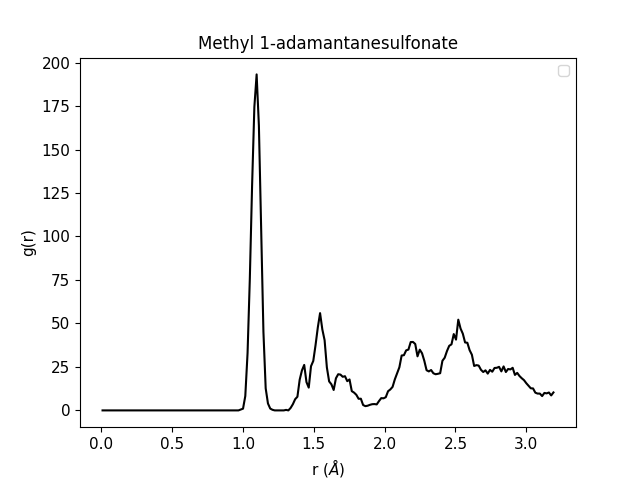}
    \includegraphics[width=.4\columnwidth]{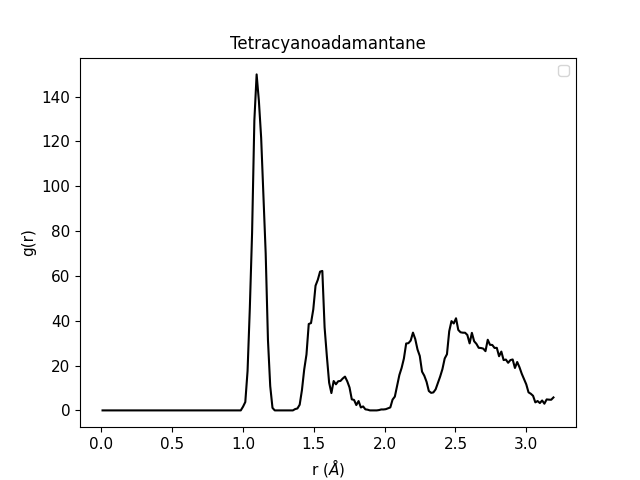}
    \caption{\label{fig:mips}  Radial Distribution Functions of the Seven Diamondoids}
	
\end{figure}

The radial distribution functions (RDFs) obtained from the molecular dynamics trajectories provide a direct measure of how each functionalized diamondoid molecule arranges around others of its kind when self-assembling into larger aggregates. By comparing the positions, intensities, and number of pronounced peaks in the RDF plots, we can infer differences in intermolecular interactions, local packing efficiencies, and the extent of structural order each functional group imparts.
For 1,3,5,7-Tetraaminoadamantane, the RDF exhibits a sharply defined and intense first peak at approximately 1.0 \AA, indicative of strong and localized short-range order. The presence of amino substituents fosters hydrogen-bonding-like interactions and directional dipole-dipole contacts, which likely reinforce a tightly packed first coordination shell. Subsequent peaks, though diminished in magnitude, remain well-defined, suggesting that while the immediate neighbor shell is robustly structured, the outer shells are moderately ordered, sustained by a combination of hydrogen bonding and steric fit within the aggregate \cite{33,34}.
In contrast, 1,3,5,7-Tetrahydroxyadamantane shows similarly intense initial peaks but with a subtly broader distribution. The hydroxyl groups readily form hydrogen bonds with neighboring molecules, promoting a stable and somewhat more expanded first coordination shell due to the possibility of directional O–H•••O interactions. Although the first peak remains distinct, the subsequent peaks are slightly less pronounced than in the tetraamino case, reflecting a balance between strong hydrogen bonds and steric constraints that yield a more fluid but still structured surrounding environment \cite{35, 36}.
For 1,3,5,7-Tetraphenyladamantane, the RDF pattern shifts to a profile where the first peak remains intense but subsequent peaks are dampened and broadened. Phenyl groups introduce increased steric bulk and potential $\pi$-$\pi$ stacking interactions. However, these interactions are typically weaker and more spatially diffuse than hydrogen bonds. The net result is a first coordination shell dominated by spatial packing of large substituents, followed by less distinct secondary shells. The phenyl groups likely disrupt the ability of molecules to uniformly pack at longer ranges, leading to a more amorphous arrangement \cite{37, 38}.

The RDF of 2,2-Dimethoxyadamantane reveals a dominant, sharp first peak followed by a series of less intense, broader maxima. Methoxy substituents are polar but less capable of forming strong hydrogen-bond networks compared to amino or hydroxyl groups. The result is a close-packed primary shell—driven more by van der Waals attractions than by strong directional interactions—and a more diffuse, less ordered arrangement at longer distances. This distribution suggests that while initial molecular contact is well-defined, the subsequent layers lack strong cohesive forces to maintain pronounced ordering.\cite{39}.
The 2,8,9-Trioxa-1-phosphadamantane RDF exhibits features indicative of both polar and sterically driven packing. The first peak is narrow and intense, reflecting a well-defined primary coordination environment. The presence of oxygen and phosphorus atoms introduces partial charges and directional dipole-type interactions that yield moderately stable secondary structures. Yet, the outer shells appear somewhat more diffuse than in the amino- or hydroxy-substituted systems, suggesting that while polarity and coordination preferences exist, they are not as consistently maintained across multiple neighbor shells \cite{40}.
For Methyl 1-adamantanelsulfonate, the initial peak is similarly sharp, and subsequent peaks reveal intermediate ordering. The sulfonate group provides a strong polar anchor, potentially enabling ionic-like interactions or strong dipole stabilization. This leads to a clearly defined primary shell. The secondary peaks, though less intense, still show recognizable structure, indicating that the charged or highly polar substituents can maintain some degree of longer-range ordering. However, the increasing complexity of packing around the bulky, polar substituents yields a more irregular and extended arrangement than observed in simpler hydrogen-bonded systems \cite{41}.

Finally, Tetracyanoadamantane displays a well-defined and intense first peak, suggesting that the strong dipole moments of cyano groups encourage tight initial packing. The subsequent peaks are moderately pronounced, reflecting the ability of nitrile functionalities to engage in dipole-dipole interactions and potentially weak hydrogen bonding if residual protons are present elsewhere in the environment. This combination results in a layered arrangement where the primary shell is tightly confined and the secondary shells, while less intense, still maintain a measure of structured ordering due to directional polar interactions \cite{42}.
In summary, the RDF analyses highlight how different functional groups on the diamondoid core modulate intermolecular contacts and assembly. Strongly hydrogen-bonding substituents (amino, hydroxy) produce distinctive near-neighbor shells and more pronounced structural ordering. Bulkier groups (phenyl) and substituents that do not form stable directional hydrogen bonds (methoxy) yield sharper first peaks but diminish longer-range order. Polar and charged groups (phosphoryl, sulfonate, cyano) strike a balance between local structuring and intermediate-range organization. Overall, these observations underscore the critical role of substituent identity in shaping the self-assembly and packing patterns of functionalized diamondoids at the molecular level.

\subsection{Thermal Stability Analysis}
\begin{figure}
	\centering
	\includegraphics[width=.4\columnwidth]{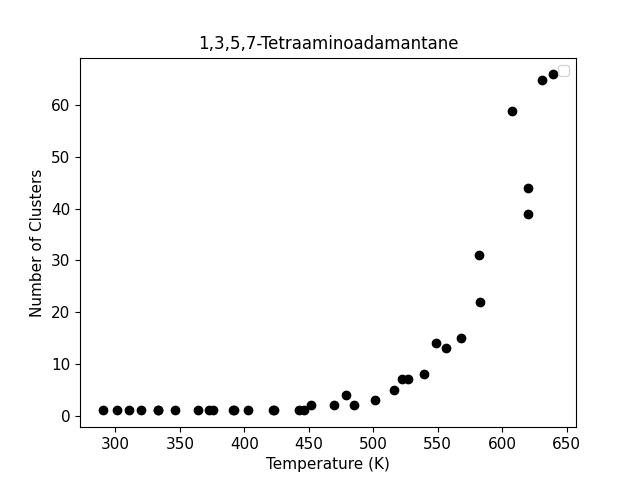}
	\includegraphics[width=.4\columnwidth]{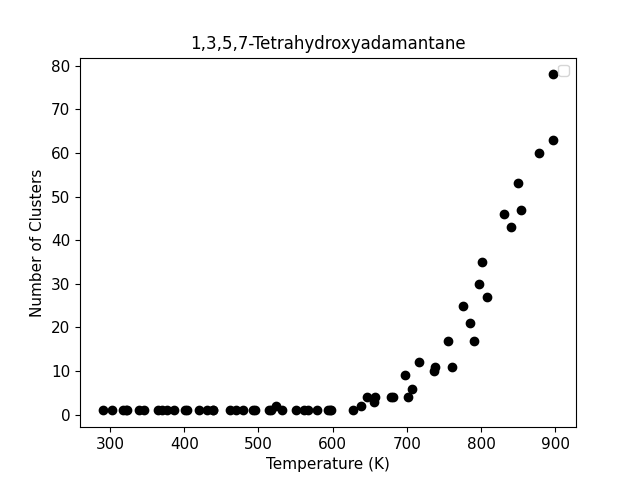}
    \includegraphics[width=.4\columnwidth]{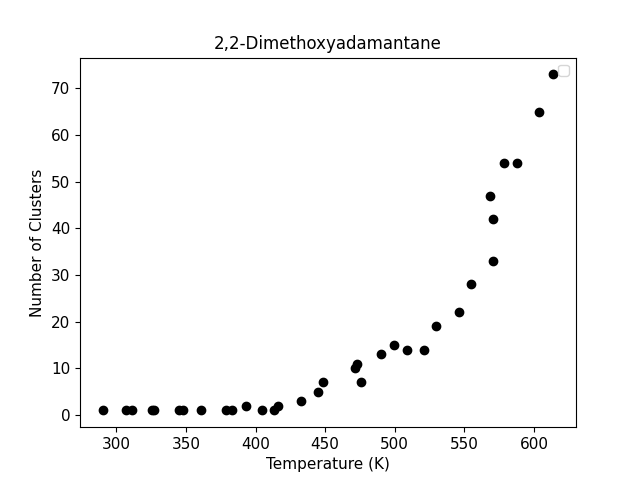}
    \includegraphics[width=.4\columnwidth]{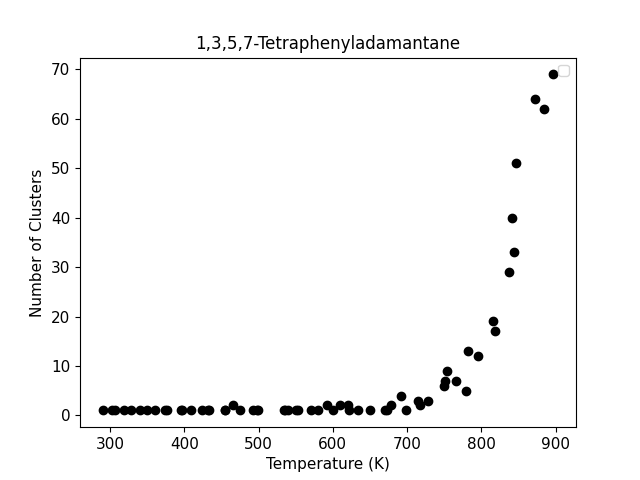}
    \includegraphics[width=.4\columnwidth]{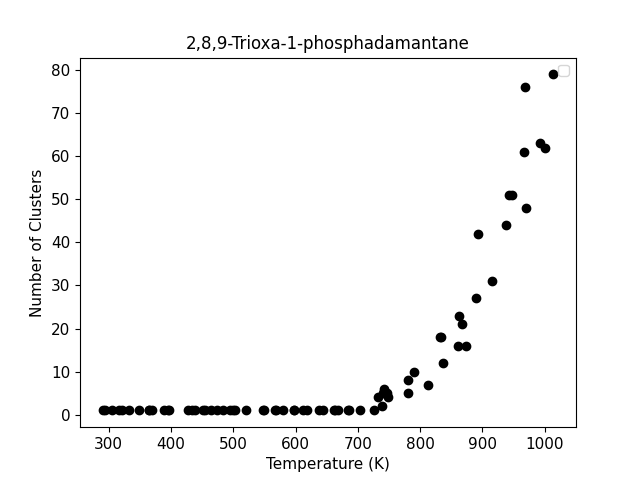}
    \includegraphics[width=.4\columnwidth]{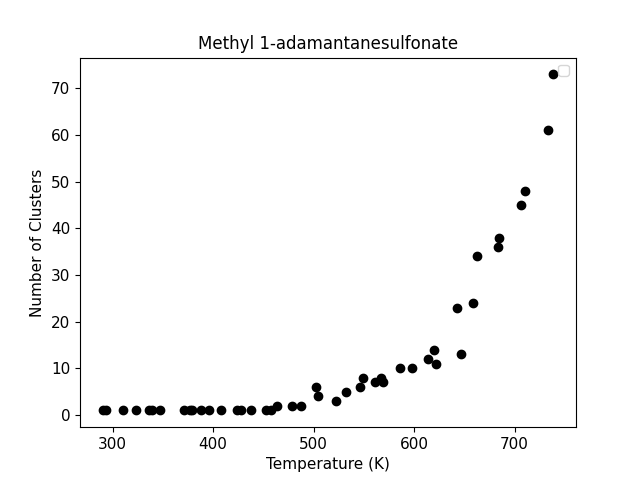}
    \includegraphics[width=.4\columnwidth]{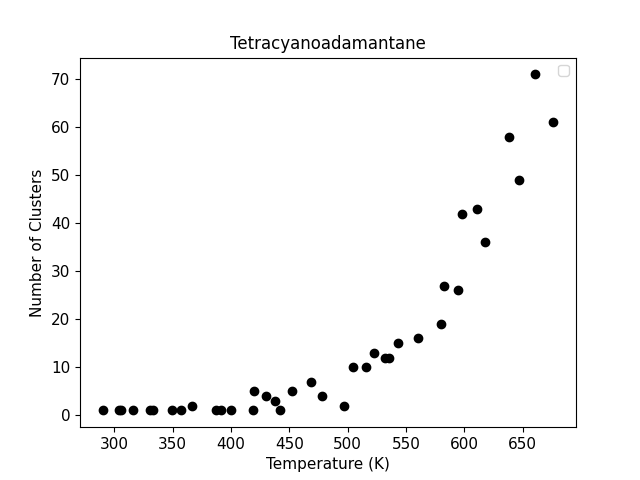}
	\caption{\label{fig:mips} Thermal Stability Plots of the Seven Diamondoids}
\end{figure}
The thermal stability of the self-assembled diamondoid superstructures, each composed of 125 functionalized diamondoid molecules, exhibits significant variation depending on the nature and position of substituents. Across all systems examined, the number of clusters—an indicator of fragmentation—remains relatively low at lower temperatures, suggesting that cohesive interactions between functionalized diamondoids are sufficient to maintain a single or a few large clusters. However, as temperature rises, thermal agitation disrupts these interactions, causing the initially well-defined assembly to break into multiple smaller clusters. The temperature at which this fragmentation becomes pronounced differs markedly among the various functional groups and substituent patterns.
For Tetracyanoadamantane, fragmentation begins at a comparatively moderate temperature. Although the assembly is stable up to around 400–450 K, a steady rise in cluster number begins soon thereafter, indicating that the nitrile functionalities do not confer strong thermal stability to the aggregate. By approximately 600 K, the assembly breaks into numerous smaller clusters, signaling that cyanide substituents, despite their polarity, do not strongly reinforce the structural cohesion under elevated thermal conditions \cite{42}.

In contrast, 2,8,9-Trioxa-1-phosphadamantane displays enhanced thermal resilience. The number of clusters remains low and nearly constant until about 600–700 K. Only beyond this high-temperature threshold does the assembly begin to fragment significantly, suggesting that the presence of phosphorus and multiple oxygen atoms creates a robust network of interactions—likely involving dipolar, hydrogen-bond-like, or coordinative interactions—that effectively resists thermal disruption. This compound’s ability to remain intact up to higher temperatures highlights the stabilizing role that heteroatomic substituents can play \cite{40}.
Methyl 1-adamantane sulfonate and 1,3,5,7-Tetrahydroxyadamantane represent intermediate cases. While they maintain structural integrity at lower temperatures, fragmentation starts to become noticeable around 450–500 K and 550-600 K respectively. The sulfonate group, with its strongly polar nature, enhances initial aggregation but cannot fully prevent eventual disassembly under thermal stress. Similarly, the hydroxyl groups in tetrahydroxyadamantane may initially form hydrogen-bonding networks, but these interactions weaken with increasing kinetic energy. Although these assemblies exhibit better stability than certain low-stability variants, they still yield to higher thermal agitation at moderately elevated temperatures.\cite{35, 36, 41}.

2,2-Dimethoxyadamantane, much like Tetracyanoadamantane, displays a relatively early onset of fragmentation. The methoxy substituents, while capable of engaging in weak dipolar interactions, do not appear to confer substantial thermal robustness. As a result, the system begins to break apart into smaller clusters at temperatures close to 400–450 K \cite{39}. By contrast, 1,3,5,7-Tetraphenyladamantane stands out for its relatively high thermal endurance, exhibiting minimal fragmentation until well above 600 K. The bulky phenyl substituents may provide steric hindrance and possible $\pi$-$\pi$ stacking or van der Waals interactions that slow down the disintegration process. Only at very high temperatures does the assembly reach a fragmentation level comparable to the other systems \cite{37, 38}.
1,3,5,7-Tetraaminoadamantane falls closer to the less stable group, beginning to show fragmentation around 400–450 K. Although amine groups can participate in hydrogen bonding and other stabilizing interactions, their arrangement in the diamondoid framework does not strongly protect the assembly against thermal agitation. As the temperature approaches 600–650 K, significant fragmentation ensues \cite{33,34}.

In summary, the results underscore a clear trend: the chemical nature and bulkiness of the substituents critically influence the thermal robustness of the diamondoid assemblies. Diamondoids substituted with more complex or bulkier functionalities (e.g., phosphadamantane, tetraphenyladamantane) remain intact up to relatively high temperatures, while those with simpler or less stabilizing groups (e.g., cyanide, methoxy, amino) fragment at lower temperatures. The interplay of hydrogen bonding, dipolar interactions, $\pi$-$\pi$ stacking, and steric effects dictates the onset of thermal-induced disassembly. Ultimately, while all systems eventually succumb to thermal agitation, the rate and temperature range over which fragmentation occurs vary significantly, revealing opportunities to tailor diamondoid functionalities for targeted thermal stability in self-assembled nanostructures.

\begin{table}[h!]
    \centering
    \caption{Radius of Gyration of Functionalized Diamondoid Assemblies}
    \label{tab:thermal}
    \begin{tabular}{@{}lll@{}}
        \toprule
        \textbf{Compound} & \textbf{Radius of Gyration ($R_g$, \AA)} \\ \midrule
        1,3,5,7-Tetraaminoadamantane      & 20.8404 \\
        1,3,5,7-Tetrahydroxyadamantane     & 22.6701 \\
        1,3,5,7-Tetraphenyladamantane      & 28.5014 \\
        2,2-Dimethoxyadamantane            & 17.7154 \\
        2,8,9-Trioxa-1-phosphadamantane    & 16.7982 \\
        Methyl 1-adamantanesulfonate        &  20.0798 \\
        Tetracyanoadamantane                &  19.4295 \\ \bottomrule
    \end{tabular}
    \vspace{0.2cm}
   
\end{table}

\subsection{Radius of Gyration Analysis}
The self-assembly simulations of the functionalized diamondoids produced a range of final aggregated structures with distinct spatial arrangements, as quantified by the radius of gyration (Rg). The Rg values, measured after the assembly of 125 identical diamondoid units, varied notably depending on the nature of the substituents. These differences highlight how the balance between steric bulk, polarity, hydrogen-bonding capacity, and overall electronic effects can influence the nanoscale packing and morphology of the assembled clusters. (Table~\ref{tab:thermal}).
Among the diamondoids examined, 1,3,5,7-Tetraphenyladamantane exhibited the largest radius of gyration (28.5014 \AA). The pronounced bulkiness and rigidity of the phenyl groups presumably impede close packing and create more open, less dense aggregates. These phenyl substituents, due to their inability to form strong directional hydrogen bonds and their comparatively large steric volume, tend to lead to looser assemblies. As a result, the cluster extends further into space, producing a high Rg value \cite{37, 38}.

In contrast, 2,8,9-Trioxa-1-phosphadamantane \cite{40}, with its Rg of 16.7982 \AA, showed the most compact aggregation. The introduction of heteroatoms, such as oxygen and phosphorus, can enhance polar interactions and potentially coordinate intermolecular hydrogen bonding with other molecules in the cluster, promoting tighter packing. These interactions may lead to more efficient space filling and a reduction in the overall structural footprint of the final assembly. Similarly, 2,2-Dimethoxyadamantane \cite{39} (17.7154 \AA) formed relatively compact structures, indicating that methoxy substituents, while introducing polarity and some steric effects, still allowed for a more efficient packing than more sterically demanding or less polar groups.
Functional groups that are capable of hydrogen bonding, such as amine and hydroxyl groups, can create well-organized interactions that influence cluster size. 1,3,5,7-Tetrahydroxyadamantane \cite{35, 36}, for example, produced an assembly with an Rg of 22.6701 \AA. Although hydrogen bonding typically promotes more compact packing, in this case the arrangement may have favored extended networks of hydrogen-bonded chains or branching motifs, preventing the cluster from adopting the most space-efficient geometry. Similarly, 1,3,5,7-Tetraaminoadamantane \cite{33, 34}, with an Rg of 20.8404 \AA, suggests a delicate interplay between hydrogen bonding and steric considerations, as the amino groups can form directional interactions but still maintain a relatively open arrangement compared to the more compact methoxy- and oxa-phosphadamantane-based assemblies.

Substituents with strong electron-withdrawing properties, such as the sulfonate and cyano groups, can also influence cluster formation. Methyl 1-adamantanesulfonate \cite{41} with an Rg of 20.0798 \AA and Tetracyanoadamantane \cite{42} with an Rg of 19.4295 \AA both formed intermediate-sized aggregates. The sulfonate group’s polarity and potential for ionic interactions could promote partial clustering into subdomains within the assembly, while the cyano groups, although polar, are not as bulky and may allow for somewhat more efficient stacking. However, the capacity of these groups to engage in directional dipole-dipole interactions without significant steric encumbrance likely prevented the formation of very loose aggregates and contributed to moderately compact assemblies.
Overall, the results indicate that subtle changes in substituent size, shape, and functionality have a pronounced effect on the self-assembled morphology of diamondoid-based clusters. Bulky substituents tend to produce more expansive assemblies, while polar and hydrogen-bonding groups can either tighten or loosen the packing, depending on the balance of interactions. These findings underscore the importance of functional group selection in tuning the self-assembly properties of diamondoid materials at the nanoscale, offering insights into how molecular design principles can direct the construction of tailored nanostructures.

\section{Conclusion}
In this study, we have systematically examined how different functional groups modulate the self-assembly behavior of a series of functionalized diamondoids. By performing molecular dynamics simulations on seven distinct substituted adamantane derivatives, we elucidated the interplay of intermolecular interactions, thermal stability, and structural compactness within these nanoscale aggregates. The results highlight that even subtle variations in substituent identity—ranging from amino, hydroxy, methoxy, phenyl, phospho-oxa, sulfonate, to cyano groups—profoundly influence the organization, thermal endurance, and overall morphology of the self-assembled clusters.

The radial distribution function analyses underscored the key role of hydrogen bonding and polarity in establishing early-stage order. Strong hydrogen-bond donors and acceptors (e.g., amino and hydroxyl groups) fostered well-defined first coordination shells and more pronounced structural ordering, while bulky or less interactive substituents (e.g., phenyl, methoxy) resulted in looser, more amorphous assemblies. Likewise, polar and charged groups (such as sulfonate or phosphoramidates) struck a balance between local structuring and intermediate-range organization, demonstrating that appropriately chosen substituents can maintain structural coherence beyond the nearest neighbors.

Thermal stability assessments revealed that substituent chemistry significantly affects the temperature at which assemblies fragment. Assemblies containing complex, bulky, or heteroatom-rich functional groups (e.g., tetraphenyladamantane, trioxa-phosphadamantane) remained intact at higher temperatures, benefiting from diverse noncovalent interactions—such as dipole-dipole contacts, $\pi$-$\pi$ stacking, and steric hindrance—that slow down thermal dissociation. Conversely, simpler functional groups conferring weaker directional interactions—such as cyanides or methoxy substituents—led to a lower resistance to thermal disruption. This finding offers valuable design criteria for engineering diamondoid-based materials with tailored thermal robustness.

The radius of gyration analyses further emphasized how substituent size, polarity, and hydrogen-bonding capabilities can tune the compactness of the self-assembled aggregates. Larger, more rigid substituents produced expansive clusters, while strong polar or hydrogen-bonding functionalities could either tighten or loosen packing, depending on the intricate balance of interactions. This detailed structural insight informs future strategies for constructing nanomaterials where molecular architecture is finely adjusted to achieve desired nanoscale morphologies and properties.

In summary, our findings provide a foundational understanding of how functionalization modulates the self-assembly characteristics, thermal stability, and structural compactness of diamondoid-based nanomaterials. These insights are directly relevant to the rational design and fabrication of advanced materials for applications in nanotechnology, pharmaceuticals, and electronics, where precise molecular control can yield structures with optimized mechanical strength, thermal endurance, and functional performance. Future work could involve studying solvent effects, mixed functionalization, or larger-scale assemblies, aiming to further refine the connection between chemical substitution patterns and emergent bulk properties. Ultimately, this research contributes valuable design principles for exploiting the rich chemical versatility of diamondoids in the creation of next-generation functional nanomaterials.

\end{document}